\documentstyle[11pt,fleqn,epsf]{article}
\catcode`\"=\active\let"=\"
\def\3{\ss }
\def\thebibliography#1{\subsection*{\refname}\small\list
 {[\arabic{enumi}]}{\settowidth\labelwidth{[#1]}\leftmargin\labelwidth
 \advance\leftmargin\labelsep
\setlength{\itemsep}{0pt}
\setlength{\parsep}{0pt}
 \usecounter{enumi}}
 \def\newblock{\hskip .11em plus .33em minus -.07em}
 \sloppy
 \sfcode`\.=1000\relax}

\def\refname{\hspace*{2cm}References} 
\def\ga{\mathrel{\mathchoice {\vcenter{\offinterlineskip\halign{\hfil
$\displaystyle##$\hfil\cr>\cr\sim\cr}}}
{\vcenter{\offinterlineskip\halign{\hfil$\textstyle##$\hfil\cr
>\cr\sim\cr}}}
{\vcenter{\offinterlineskip\halign{\hfil$\scriptstyle##$\hfil\cr
>\cr\sim\cr}}}
{\vcenter{\offinterlineskip\halign{\hfil$\scriptscriptstyle##$\hfil\cr
>\cr\sim\cr}}}}}
\newcommand\etal{et al.\,}

\begin{document}
\noindent
\hspace*{2cm} {\Large\bf{A Semi-relativistic Equation of State \\ 
\hspace*{2cm}            for Stellar Interiors}} \\ \\
\hspace*{2cm} {\sc W. Stolzmann} (a,b), 
              {\sc T. Bl\"ocker} (c)  \\ \\
\hspace*{2cm} (a) Institut f\"ur Theoretische Physik und Astrophysik, Kiel, 
                  Germany \\
\hspace*{2cm} (b) Astrophysikalisches Institut Potsdam, Germany \\
\hspace*{2cm} (c) Max-Planck-Institut f\"ur Radioastronomie, Bonn, Germany
\subsection*{\hspace*{2cm} Abstract}
{\small
\noindent 
Using the technique of Pad\'e Approximants for the correlation
contributions of charged particles we are able to
reproduce complex mathematical expressions by simple algebraic formulas.
The Pad\'e formulas are analytical expressions, which interpolate
between certain density-temperature regions and are characterized
by exact asymptotics. We present a semi-relativistic description
of the thermodynamics applicable for stellar interiors.
Comparisons with equation-of-state data obtained by other calculational 
schemes are presented.
}
\subsection*{\hspace*{2cm} 1. Introduction}
The equation of state (EOS) formalism for astrophysical applications
requires an accurate calculation of the non-ideal effects
in plasmas over a wide range of densities and temperatures.
Many successful efforts have been made to provide EOS-tables
\cite{fograhor77,mama79,iit87,mhd88,effgp91,scvh95,ris96}
which are applicable to describe thermodynamical properties 
of stellar matter.
However, detailed comparisons of quantities composed of
second-order derivatives of the free Helmholtz energy 
(e.g.\ specific heat, adiabatic temperature gradient)
reveal considerable discontinuities (see, e.g., \cite{scvh95}).
We present an EOS-formalism which avoids such numerical deviations 
and is well suited for example to evaluate the thermodynamics of 
stellar interiors.
\subsection*{\hspace{2cm} 2. Thermodynamical Functions}
We start with the thermodymamical description of a fully ionized plasma
\begin{equation}  \label{Eicsredu}
\Sigma  =  N_{\rm e}kT \left[\sigma_{\rm e }^{\rm id}
                           + \sigma_{\rm ee}^{\rm x }
                           + \sigma_{\rm ee}^{\rm c } \right]  +
           N_{\rm i}kT \left[\sigma_{\rm i }^{\rm id}
                           + \sigma_{\rm ii}^{\rm c }
                           + \sigma_{\rm ii}^{\rm cq}
                           + \sigma_{\rm ie}^{\rm c } \right] \;\; ,
\end{equation}
where
$\Sigma=\{ F, G, P V, U, V/K_{\rm T}, \Phi_{\rm S} V,
C_{\rm V} T \}$
and $\sigma=\{ f, g, p, u, 1/k_{\rm T}, \phi_{\rm S}, c_{\rm V} \}$
symbolize the various thermodynamical functions, as the
free Helmholtz energy $F$, free Gibbs energy $G$, pressure $P$,
internal energy $U$, inverse isothermal compressibility $1/K_{\rm T}$,   
strain coefficient $\Phi_{\rm S}$, and isochoric specific heat $C_{\rm V}$.
Quantities given by small letters are the corresponding
potentials normalized to $N_{\rm e,(i)} k_{\rm B}T$.  
Terms labeled by id, x, and c denote contributions due to
ideality, exchange, and correlation contributions of  the plasma species
$a={\rm e},{\rm i}$.
The term labeled by cq takes into account quantum effects of ions. \\
Furthermore we have to determine the isobaric specific heat $C_{\rm P}$
by the relation \cite{cogiu68}
\begin{equation}    \label{Ecpcv}
\gamma = \frac{C_{\rm P}}{C_{\rm V}}   
       = 1 + \frac{V}{T} \frac{K_{\rm T}} {C_{\rm V}}
                            \Phi_{\rm S}^{2} \; , \; 
\quad
\frac {1}{K_{\rm T}} = P \chi_{\rho} = - V \left( \frac{\partial P}
                    {\partial V} \right)_{T} \; , \;
\quad 
\Phi_{\rm S} = P \chi_{\rm T} =  T \left( \frac{\partial P}
                                           {\partial T} \right)_{V} \; .
\end{equation}
$\chi_{\rho}$ and $\chi_{\rm T}$ are the density and temperature exponents
in the EOS-formalism.
The adiabatic temperature gradient defined by
$\nabla_{\rm ad} = (\partial \ln T/\partial \ln P)_{S}$
($S$ is the entropy) is given by \cite{cogiu68,ris96}
\begin{equation}    \label{Eadgra}
\nabla_{\rm ad} = \frac{P V}{C_{\rm P}T}\frac{\chi_{\rm T}}{\chi_{\rho}}  
                = \frac{P V}{C_{\rm P}T} K_{\rm T} {\Phi_{\rm S}}
                = \frac{\Gamma_{2}-1}{\Gamma_{2}}
                = \frac{\Gamma_{3}-1}{\Gamma_{1}}  \;\; ,
\end{equation}
with the adiabatic exponents $\Gamma_{1}=\gamma/P K_{\rm T}, \;
\Gamma_{2}=\gamma/(\gamma - P K_{\rm T}\gamma_{\rm G})$ and
$\Gamma_{3}=1 + \gamma_{\rm G}$,
and the generalized Gr\"uneisen coefficient 
$\gamma_{\rm G}=V \Phi_{\rm S}/T C_{\rm V}$.

The ideal and exchange contributions, 
$\sigma_{\rm e}^{\rm id}$ and $\sigma_{\rm ee}^{\rm x}$, 
are calculated in a relativistic approach, 
which covers the two approximations: 
{\bf a})  arbitrary degeneracy $(\psi)$  
          and weak relativity $(\lambda < 1)$ and
{\bf b})  strong degeneracy    $(\psi > 1)$ and arbitrary
          relativity $(\lambda=kT/mc^2)$. 
Details are given for free Helmholtz energy, free Gibbs energy, 
and pressure in \cite{sb96aua}.

The correlation terms $\sigma_{\rm ee}^{\rm c}$, $\sigma_{\rm ii}^{\rm c}$,
and $\sigma_{\rm ie }^{\rm c}$ are taken into account by Pad\'e approximants
which are modified versions of those given in \cite{sb96aua}.
The new Pad\'e approximants are determined by a rearrangement of the 
quantum virial function which is dominant in the weakly coupled regime.
Moreover the Pad\'e formula for the ion-electron subsystem is extented by 
linear $\Gamma$-asymptotics for the classical region \cite{steb98}. 
The ionic quantum correction $\sigma_{\rm ii}^{\rm cq}$ are described by 
\cite{nnn87}.

In this paper we present our new Pad\'{e} approximant for the 
ion-ion correlation. 
Starting with the free energy the complete set of the quantities summarized 
by $\sigma_{\rm ii}^{\rm c}$ will be given. We mention that the Pad\'e 
formulas for the quantities related by the first and second order derivatives 
of the free energy are constructed from the analytical limiting laws, which 
are known to be asymptotically exact. The intermediate region will be fitted 
by new parameters. Of course, the quality of this procedure with regard to  
the thermodynamic consistency is determined by comparisons with the 
numerical data produced from corresponding theories.
The advantage is that the Pad\'e approximants for our set of thermodynamic 
functions are expressed by their correct asymptotics, which cover a wide 
parameter-region.          

For the ionic subsystem we use the new formulas for the free Helmholtz-
and Gibbs energy with 
$\Gamma_{\rm i}=\langle Z^{5/3} \rangle e^{2} (4 \pi n_{\rm e}/3)^{1/3}
/k_{\rm B}T$, 
\begin{equation}  \label{Efiic}
 f_{\rm ii}^{\rm c}= - \frac{b_{0} \Gamma_{\rm i}^{3/2}
                  \left[ 1 + b_{3} \Gamma_{\rm i}^{3/2}
                        \left( \ln \Gamma_{\rm i} + B_{0} \right) \right]
    +  b_{2} \Gamma_{\rm i}^{6} \varepsilon_{\rm ii}(\Gamma_{\rm i})}
                         {1 - b_{1} \Gamma_{\rm i}^{3} 
                        \left( \ln \Gamma_{\rm i} + B_{1} \right)
    +  b_{2} \Gamma_{\rm i}^{6}}  
\end{equation}
\begin{equation}  \label{Egiic}
 g_{\rm ii}^{\rm c} =-\frac{t_{0} \Gamma_{\rm i}^{3/2}
                \left[ 1 +  t_{3} \Gamma_{\rm i}^{3/2}
                \left( \ln \Gamma_{\rm i} + B_{0} + \frac{1}{6} \right) \right] 
   +  t_{2} \Gamma_{\rm i}^{6} \mu_{\rm ii}(\Gamma_{\rm i})}
                                      {1 - t_{1} \Gamma_{\rm i}^{3}
                \left(\ln \Gamma_{\rm i} + B_{1} + \frac{2}{15} \right )
                \left[ 1 + \frac{3}{4} \Gamma_{\rm i}^{3/2} \right]
   +  t_{2} \Gamma_{\rm i}^{6}}
\end{equation}
with
\begin{equation}  
B_{0} = \frac{2}{3}\left(2C_{\rm E}+\frac{3}{2}\ln 3-\frac{11}{6}\right) \; ,
\quad\quad
B_{1} = \frac{2}{3}\left(2C_{\rm E}+\frac{1}{2}\ln 3+2\ln 2-\frac{17}{6}\right)
        -0.4765
\end{equation}
and the coefficients
$b_{0} = 1 / \sqrt{3} \; ,
b_{1} = 3 \sqrt{3}/8b_{0}, \; 
b_{2} = 100,                       \;
b_{3} = b_{1}/\sqrt{3}$, 
and $t_{0} =3 b_{0}/2 , \; t_{1} = 5 b_{1}/3 , \;
     t_{2} =  b_{2}   , \; t_{3} = 4 b_{3}/3$  .
Eq.~(\ref{Efiic}) is based on the classical $\Gamma_{\rm i} < 1$ result from 
Cohen and Murphy \cite{comu69} and for $\Gamma_{\rm i} \ga 1$ we take into 
account the Monte-Carlo fits for the free energy 
$\varepsilon_{\rm ii} (\Gamma_{\rm i})$ of the liquid and solid one-component
plasma (OCP) from Stringfellow \etal \cite{sds90}. 

The internal energy, the compressibility, and the strain coefficient 
are determined by \cite{han73}
\begin{equation}   \label{Euiicpad}
u_{\rm ii}^{\rm c}  = 3 \;     p_{\rm ii}^{\rm c}
                    = 3 \left( g_{\rm ii}^{\rm c}
                              -f_{\rm ii}^{\rm c} \right)   \;\;, 
\end{equation}
\begin{equation}
\frac{1}{k_{\rm Tii}^{\rm c}} = - \frac{1}{9} \; c_{\rm Vii}^{\rm c}
                                + \frac{4}{9} \; u_{\rm ii }^{\rm c} \;\;,
\quad\quad\quad
%
\phi_{\rm Sii}^{\rm c} = \frac{1}{3} \; c_{\rm Vii}^{\rm c}  \;\;.
\end{equation}
Note, that a factor 1/3 is missing in 
the second term of $1/k_{\rm Tii}^{\rm c}$ given by Hansen \cite{han73}.

For the isochoric specific heat we apply the Pad\'e approximant
\begin{equation}     \label{Ecviicpad}
c_{\rm Vii}^{\rm c} =
   \frac{q_{0} \Gamma_{\rm i}^{3/2}  \left[ 1 + q_{3} \Gamma_{\rm i}^{3/2}
       \left(\ln \Gamma_{\rm i}+B_{0} + \frac{5}{6} \right ) \right]
 +       q_{2} \Gamma_{\rm i}^{6} \varrho_{\rm ii}(\Gamma_{\rm i})}   
   { 1 - q_{1} \Gamma_{\rm i}^{3}
       \left(\ln \Gamma_{\rm i}+B_{1} + \frac{32}{63} \right )
 +       q_{2} \Gamma_{\rm i}^{6}}
\end{equation}
with the coefficients 
$q_{0} = 3 b_{0}/4 , \; q_{1} = 7 b_{1} , \;  
 q_{2} = 5 b_{2}   , \; q_{3} = 8 b_{3}$.
\begin{equation}
\mu_{\rm ii}(\Gamma_{\rm i}) = \varepsilon_{\rm ii}(\Gamma_{\rm i}) +
                               \frac{1}{3} \Gamma_{\rm i}
 \frac{\rm d \varepsilon_{\rm ii} (\Gamma_{\rm i})}{\rm d \Gamma_{\rm i}} 
\;\; , \quad\quad\quad 
\varrho_{\rm ii}(\Gamma_{\rm i}) = \Gamma_{\rm i}^{2}
    \frac{{\rm d}^{2} \varepsilon_{\rm ii}(\Gamma_{\rm i})}
                                  {{\rm d} \Gamma_{\rm i}^{2}} \;\; .
\end{equation}
\subsection*{\hspace{2cm} 3. Results and Discussion}
\begin{figure}
\epsfxsize=0.47\textwidth
\mbox{\epsffile{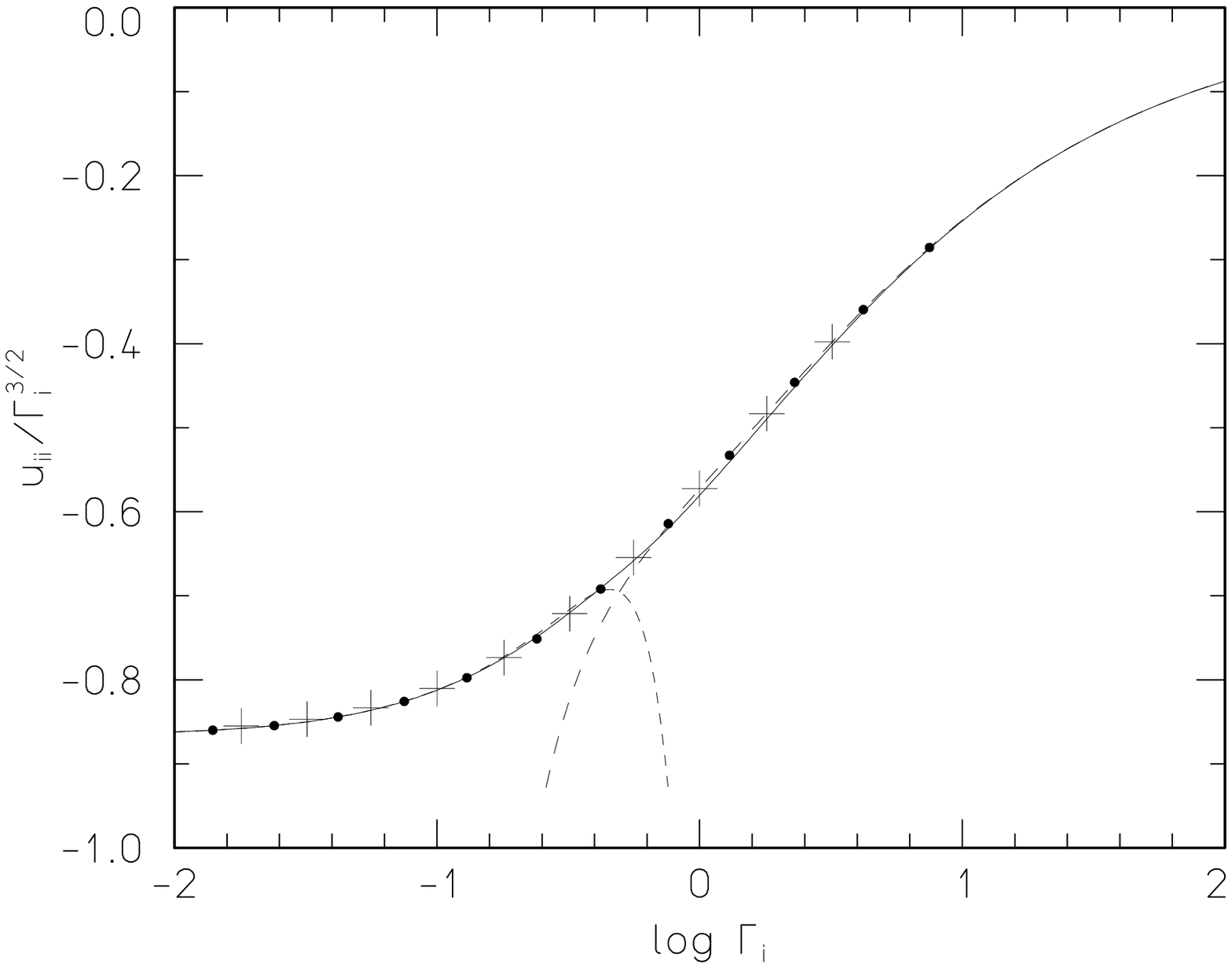}}
\hspace*{0.5cm}
\epsfxsize=0.47\textwidth
\mbox{\epsffile{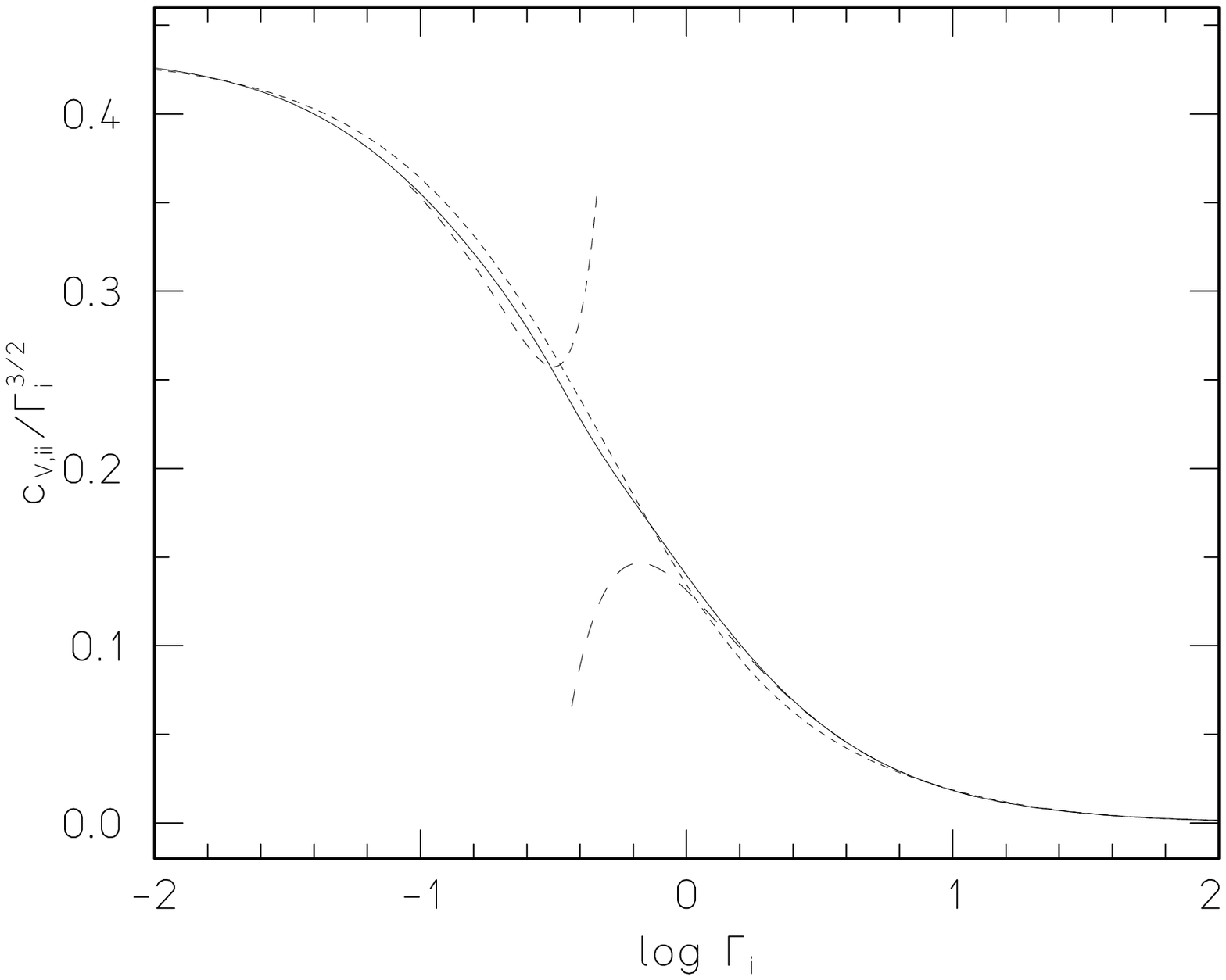}}
\hspace*{0.7cm}\parbox{14.7cm}{
\caption[u_cv_ii_ocp]
{ Internal energy ({\it left}) and isochoric specific heat ({\it right}) 
  of the OCP model. The short-dashed lines refer to the $\Gamma_{\rm i} < 1$ 
  asymptotics from Cohen and Murphy \cite{comu69} and the long-dashed line 
  to the $\Gamma_{\rm i} \ga 1$ fit from Stringfellow \etal \cite{sds90}.
{\it Left}:  
  The solid line refers to Eq.~(\ref{Euiicpad}), the dots
  to Kahlbaum \cite{toka96}, and the crosses to Chabrier and Potekhin 
  \cite{chpo98}.
{\it Right}:
  The solid line refers to Eq.~(\ref{Ecviicpad}) and the dotted one to 
  Hansen \cite{han73}.
}       \label{F_u_cv_ii_ocp}
}
\end{figure}  
\begin{figure}
\epsfxsize=0.47\textwidth   
\mbox{\epsffile{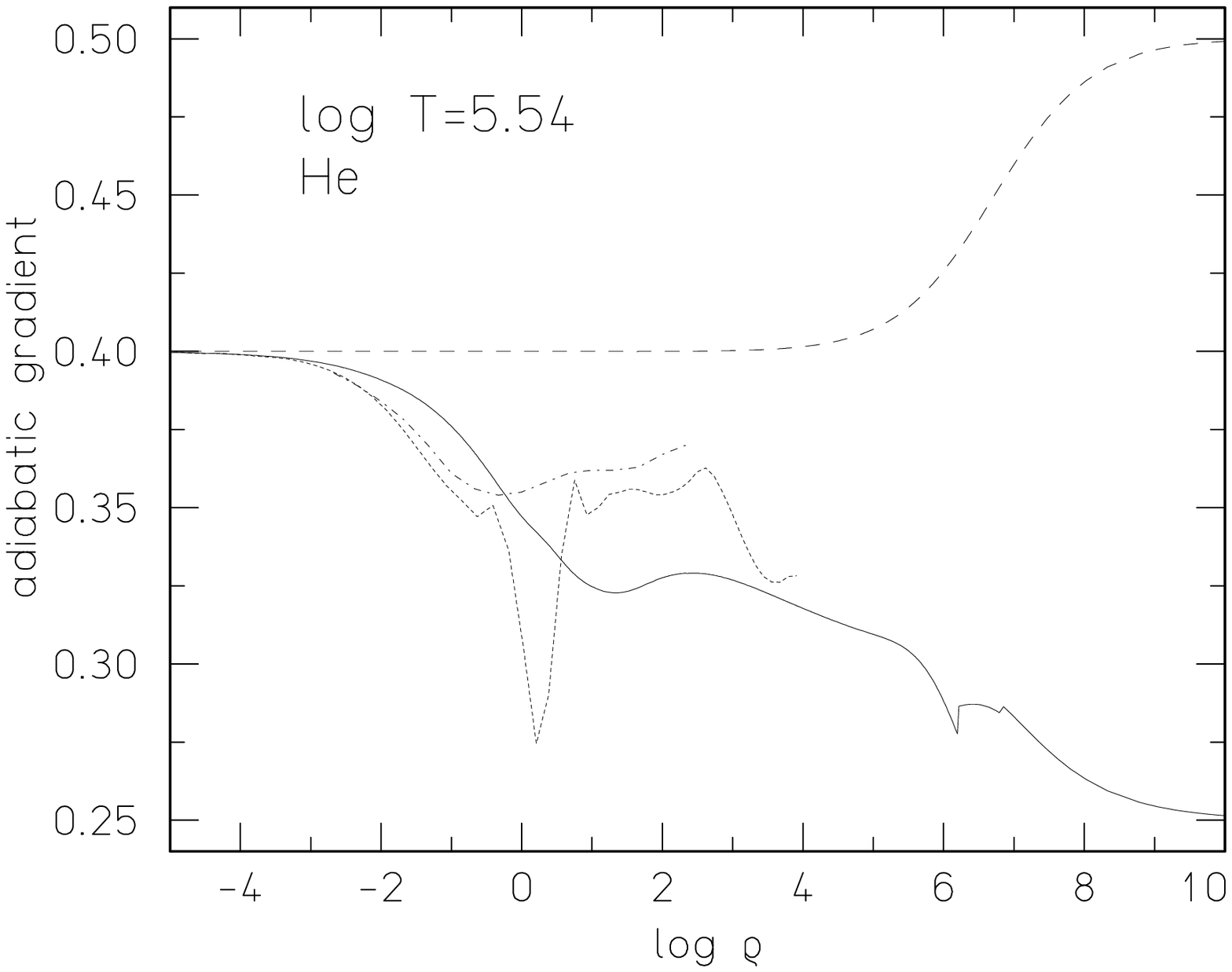}}
\hspace*{0.5cm}
\epsfxsize=0.47\textwidth
\mbox{\epsffile{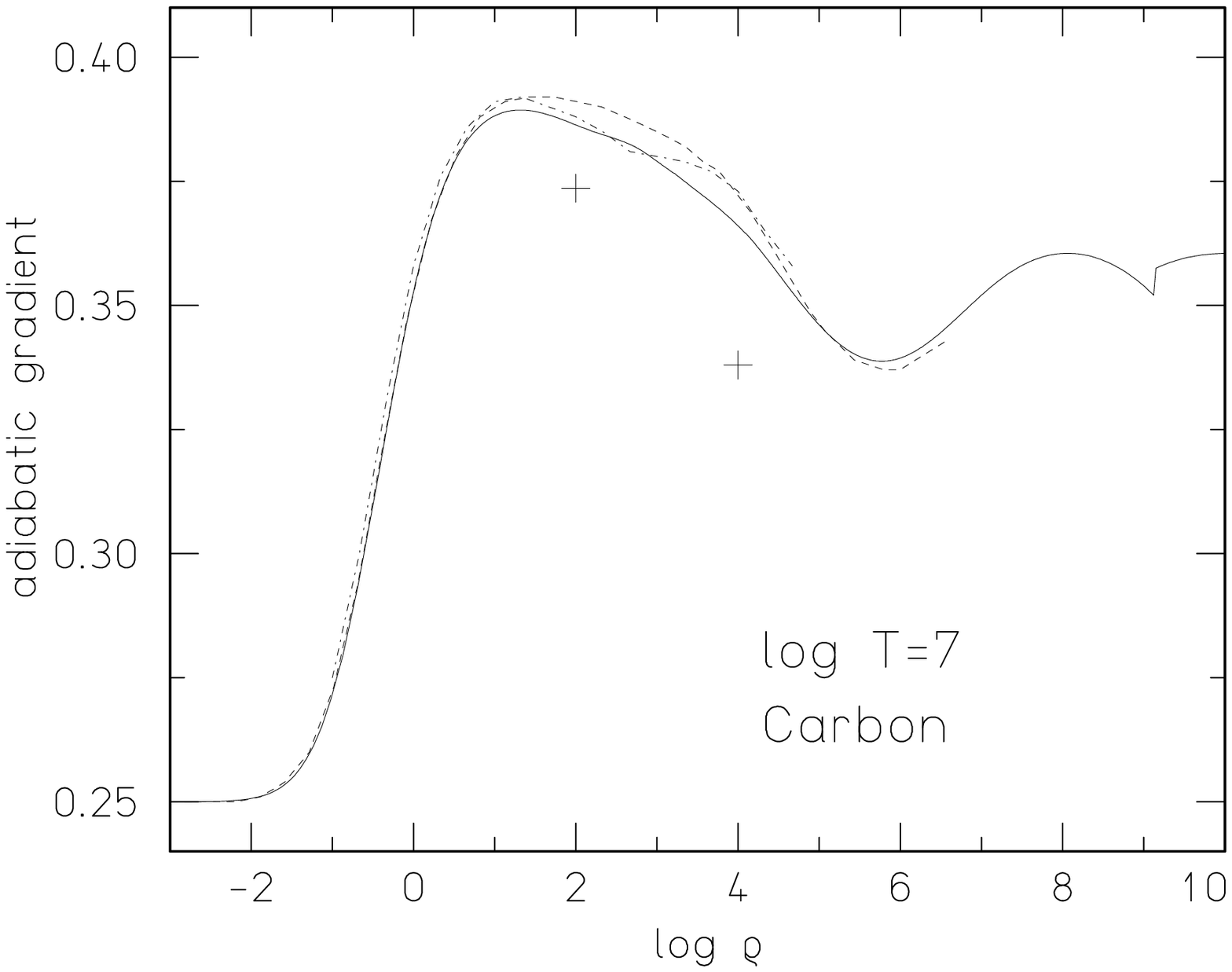}}
\hspace*{0.7cm}\parbox{14.7cm}{
\caption[nab_554_He_id]
{Adiabatic temperature gradient $\nabla_{\rm ad}$ vs.\ density for helium 
 at $T=10^{5.54}$\,K without radiation ({\it left}) and for carbon at 
 $T=10^{7}$\,K with radiation ({\it right}). 
 Note, that the discontinuities at $\log \rho \approx 6$ ({\it left}) and 
 $\log \rho \approx 9$ ({\it right}) are caused by the fluid-solid phase 
 transition at $\Gamma=178$ \cite{sds90}.
 The solid curves refer to Eqs.~(\ref{Eadgra}), (\ref{Ecpcv}), (\ref{Eicsredu})
 and the dashed-dotted to \cite{fograhor77}. 
{\it Left}:
 The dotted line refers to \cite{scvh95}. 
 For comparision is shown the ideality (long-dashed line).
{\it Right}:
 The dashed line refers to \cite{stran88} and the crosses to \cite{lamb74}.
}          \label{Fnab_7_C} 
}
\end{figure}
In Fig.~\ref{F_u_cv_ii_ocp} our new Pad\'e approximants
for the internal energy (\ref{Euiicpad}) and the isochoric specific heat 
(\ref{Ecviicpad}) are compared with other closed-form parametrizations 
\cite{han73,toka96,chpo98}. 
Obviously, the internal energy is already covered by the asymptotic 
theories given in \cite{comu69,sds90} over a wide $\Gamma_{\rm i}$-region. 
A detailed comparison between \cite{toka96} and the simplified fit formula 
from Chabrier and Potekhin \cite{chpo98} was carried out by 
Potekhin \cite{aypo98} recently.
Comparisons between the internal energy derived from our earlier 
Pad\'e approximants for the free energy \cite{sb96aua,web90} and a new fit 
formula from Chabrier and Potekhin \cite{chpo98} are shown in \cite{chpo98}.    
It is noteworthy, that our former Pad\'e approximant for $f_{\rm ii}^{c}$
proposed by Ebeling \cite{web90} contains ionic quantum corrections, 
which deliver in the classical limit the result of Abe \cite{abe59}. 
The inclusion of quantum effects (see \cite{sb96aua}, the coefficient $b_{1}$ 
in Eq.~(65)) implies an additional temperature-dependence besides the pure 
$\Gamma$-dependence. Accordingly, the determination of thermal quantities 
such as internal energy, strain coefficient, or specific heat from the former 
Pad\'e formula for $f_{\rm ii}^{c}$ must be carried out carefully. 
Non-thermal derivatives such as pressure did not suffer from this drawback
as shown in \cite{sb96aua}.   
 
In our new calculational scheme all ionic quantum corrections are considered 
separately.  As shown in Fig.~(\ref{F_u_cv_ii_ocp}) 
the classical interpolation formulas are well in agreement.

Fig.~\ref{Fnab_7_C} illustrates the course of the adiabatic temperature 
gradient, $\nabla_{\rm ad}$, calculated by our new Pad\'e approximants
and gives comparisons with numerical results from other authors 
\cite{chpo98,han73,lamb74}.
$\nabla_{\rm ad}$ is a highly sensitive quantity, because it depends on first 
and second-order derivatives of the model free Helmholtz energy as given by 
Eq.~(\ref{Eadgra}). The low-density limes of $\nabla_{\rm ad}=0.25$ 
(right panel) is a consequence of the photonic contribution in contrast to
Fig.~\ref{Fnab_7_C} (left panel).

We have shown how by means of analytical expressions 
(see also \cite{sb96aua,steb98}) the thermodynamics of e.g. stellar interiors
can be described over a wide range of densities and temperatures.
 
Our analytical description leads to a smooth course of $\nabla_{\rm ad}$
and avoids any numerical noise.
Further comparisions are given in detail by \cite{sb98bos}.
\end{document}